\documentclass[aps,prl,twocolumn,showpacs,superscriptaddress,reprint]{revtex4-1}
\usepackage{graphicx}
\usepackage{natbib}
\usepackage{amsmath}
\usepackage{amssymb}
\usepackage{bm}
\usepackage{color}
\usepackage{float}

\begin{document}

\title{Unveiling the Hybridization Process in a Quantum Critical Ferromagnet by Ultrafast Optical Spectroscopy}
\author{Y. H. Pei}
\affiliation{State Key Laboratory of Electronic Thin Films and Integrated Devices, University of Electronic Science and Technology of China, Chengdu 611731, China}
\author{Y. J. Zhang}
\affiliation{Center for Correlated Matter and Department of Physics, Zhejiang University, Hangzhou 310058, China}
\author{Z. X. Wei}
\affiliation{State Key Laboratory of Electronic Thin Films and Integrated Devices, University of Electronic Science and Technology of China, Chengdu 611731, China}
\author{Y. X. Chen}
\affiliation{Center for Correlated Matter and Department of Physics, Zhejiang University, Hangzhou 310058, China}
\author{K. Hu}
\affiliation{State Key Laboratory of Electronic Thin Films and Integrated Devices, University of Electronic Science and Technology of China, Chengdu 611731, China}	
\author{Yi-feng Yang}
\affiliation{Beijing National Laboratory for Condensed Matter Physics, Institute of Physics, Chinese Academy of Science, Beijing 100190, China}
\affiliation{University of Chinese Academy of Sciences, Beijing 100049, China}
\affiliation{Songshan Lake Materials Laboratory, Dongguan 523808, China}
\author{H. Q. Yuan}
\email{hqyuan@zju.edu.cn}
\affiliation{Center for Correlated Matter and Department of Physics, Zhejiang University, Hangzhou 310058, China}
\affiliation{State Key Laboratory of Silicon Materials, Zhejiang University, Hangzhou 310058, China}
\affiliation{Zhejiang Province Key Laboratory of Quantum Technology and Device,	Department of Physics, Zhejiang University, Hangzhou 310058, China}
\author{J. Qi}
\email{jbqi@uestc.edu.cn}
\affiliation{State Key Laboratory of Electronic Thin Films and Integrated Devices, University of Electronic Science and Technology of China, Chengdu 611731, China}

\begin{abstract}
We report the ultrafast optical pump-probe spectroscopy measurements on the recently discovered quantum critical ferromagnet CeRh$_6$Ge$_4$. Our experimental results reveal the two-stage development of the hybridization between localized $f$ moments and conduction electrons with lowering temperature, as evidenced by (1) the presence of hybridization fluctuation for temperatures from $\sim$85 K ($T^*$) to $\sim$140 K ($T^\dagger$), and (2) the emergence of collective hybridization below the coherence temperature, $T^*$, marked by the opening of an indirect gap of 2$\Delta$ $\approx$12 meV. We also observe three coherent phonon modes being softened anomalously below $T^*$, reflecting directly their coupling with the emergent coherent heavy electrons. Our findings establish the universal nature of the hybridization process in different heavy fermion systems.
\end{abstract}
\maketitle

The ferromagnetic (FM) quantum critical point (QCP) is generally believed to be prohibited in a pure FM system as it is often interrupted by other competing phases or first-order phase transitions \cite{M. Brando2016}. The recent discovery of a FM QCP in the stoichiometric Kondo lattice compound CeRh$_6$Ge$_4$ under pressure has stimulated great interest on its origin and the nature of its associated strange metal phase \cite{B. Shen2020}. One proposal for the existence of a FM QCP is that the local moments may form a triplet resonating valence bound state in the FM state and cause a singular transformation in the patterns of their entanglement with conduction electrons as the Kondo singlets develop at the QCP \cite{Piers Coleman2020}, which leads to an abrupt jump of the Fermi surface volume. This is similar to the scenario of a local QCP in the antiferromagnetic Kondo lattice compounds \cite{Si_Science_2010}. One may naturally ask if and how different types of inter-site magnetic correlations among localized $f$ moments (even outside the quantum critical regime) might have an effect on their hybridization process with conduction electrons and the resulting Fermi surface change. It is thus important to explore the band evolution of CeRh$_6$Ge$_4$ near the Fermi energy ($E_F$) and compare it with the antiferromagnetic analogues.

In this respect, the quantum oscillations and angle-resolved photoemission spectroscopy (ARPES) are the most direct experimental approaches to probe the band structure and its evolution. However, ARPES is highly limited by its energy resolution when applied on the heavy fermion materials \cite{Q. Y. Chen2017, A. Koitzsch2008, A. Koitzsch2013}. So far, it has not been able to reveal the formation of the indirect hybridization gap with an order of meV near $E_F$. Fortunately, the ultrafast optical pump-probe spectroscopy has recently been demonstrated to provide an alternative route to detect the hybridization dynamics over a wide temperature range \cite{Y. P. Liu2020}. It can reveal not only the ``band bending" probed by the ARPES far above the coherence temperature but also the opening of the indirect hybridization gap below the coherence temperature, thus proposing a unified picture for spectroscopic and transport measurements. Although the measurements can only be performed at ambient pressure away from the quantum critical point, one may still expect some useful information on the microscopic dynamics of the hybridization physics. Indeed, this optical technique provides a unique way to investigate the dynamics of excited quasiparticles coupled to collective bosonic excitations in quantum systems \cite{J. Demsar2003, E. E. Chia2010, J. Qi2013, R. Y. Chen2016}, and thus can help us to detect simultaneously the exotic fermionic and bosonic responses near and far from the coherent temperature $T^*$, below which the heavy electron state is developed.

In this work, we report the optical pump-probe measurements on CeRh$_6$Ge$_ 4$ for the first time. Hybridization processes between the localized $f$ moments and conduction electrons are unraveled via the photoexcited quasiparticle dynamics in this material. Specifically, we have observed that the quasiparticle relaxation rate ($\gamma$) in the short timescale shows a clear reduction as the temperature decreases across $T^\dagger\approx$140 K. Remarkably, $\gamma$ exhibits an apparent fluence-dependence for $T<T^*$ (=85$\pm$5~K), in comparison with the fluence-independent behaviour for $T^*<T<T^\dagger$. Such findings enable us to predict in CeRh$_6$Ge$_4$ (1) the occurrence of a narrow indirect hybridization gap of about 12 meV in the density of states (DOS) associated with the formation of coherent heavy electrons below $T^*$, and (2) the existence of precursor hybridization fluctuations between $T^*$ and $T^\dagger$. These results are in resemblance of that reported previously for CeCoIn$_5$ \cite{Y. P. Liu2020} and independent of the low-temperature magnetic properties. By contrast, recent measurements of quantum oscillations and calculations of band structure suggest a localized nature of 4$f$ electrons in CeRh$_6$Ge$_4$ in the ground state \cite{A. Wang2021}, suggesting the necessity of a better theoretical understanding.

The ultrafast time-resolved differential reflectivity $\Delta R(t)/R$ measurements were performed on the single crystal CeRh$_6$Ge$_4$ from 5 K to room temperature using a Ti:sapphire femtosecond laser producing a pulse width of $\sim$35 fs at centre wavelength of 800 nm ($\sim$1.55 eV). Typical results are shown in Fig.~\ref{Fig1}(a). More experimental details and data can be found in the Supplemental Material. After photoexcitation, strong instantaneous rise can be clearly seen at all investigated temperatures. The lateral dynamics show obvious damped oscillations superimposed on the non-oscillating background (see Fig.~\ref{Fig1}(b)). Generally, the time evolution of $\Delta R(t)/R$ within the first tens of picoseconds can be attributed to the electron-electron (e-e) and electron-boson scattering processes in strongly correlated and metallic-like systems \cite{D. N. Basov2011, C. Gadermaier2010}. The bosons may involve phonon or other bosonic excitations \cite{C. Giannetti2016, M. C. Wang2018}.

\begin{figure}
	\includegraphics[width=8.6cm]{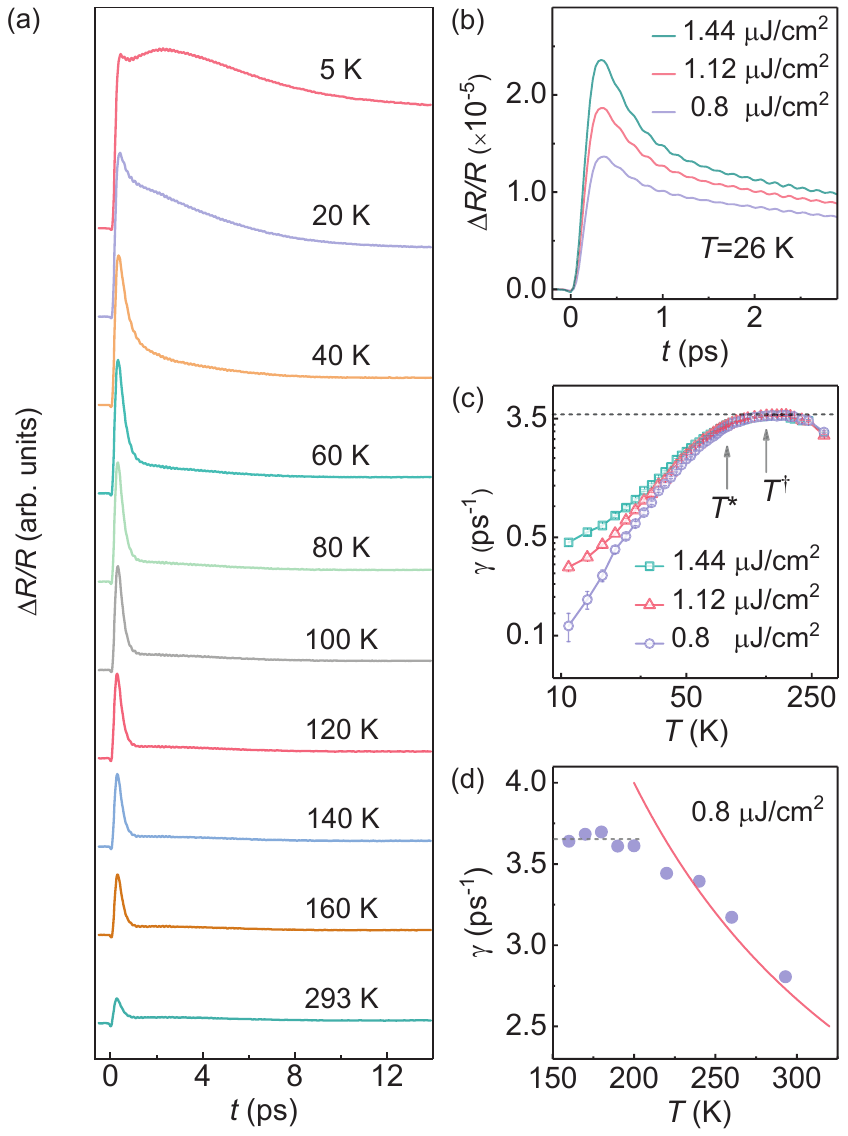}
	\caption{\label{Fig1} (a) $\Delta R(t)/R$ of CeRh$_6$Ge$_4$ as a function of temperature. (b) $\Delta R(t)/R$ at 26 K under various pump fluence. Strong fluence-dependence behaviour is clearly observed. (c) The decay rate $\gamma$ as a function of temperature for different pump fluences. Two critical temperatures ($T^*$ and $T^\dagger$) are identified. (d) $\gamma$ at high temperature regime for a fluence of 0.8~$\mu$J/cm$^2$. The red line is a fit using the nonequilibrium model described in the main text. }
	\vspace*{-0.4cm}
\end{figure}

We first focus on the non-oscillatory part of the signal. As shown in Fig.~\ref{Fig1}(b), the initial decay below $\sim1.5$~ps demonstrates a strong fluence-dependent behaviour. A bump-like behaviour consisting of a second rise clearly appears when the temperature is below $\sim$20 K, and leads to the initial decay hidden inside a tiny peak at very low temperatures. Similar signals with a second rise have also been obtained in many heavy fermion materials and other strongly correlated systems, and could originate from the excitations of electronic origins entangled with the electron-phonon (e-ph) and nonthermal e-e scatterings \cite{Y. P. Liu2020, J. Demsar2003, P. Kusar2005}. We can fit the data within $\sim$1.5 ps after the max of $\Delta R(t)/R$ signal using a single exponential formula, $\Delta R/R=Ae^{-\gamma t}$, to investigate the quasiparticle relaxation quantitatively. Here, $A$ and $\gamma$ are the amplitude and decay rate, respectively. Figure~\ref{Fig1}(c) shows the derived $\gamma$ as a function of temperature under various pump fluence. An evident fluence-dependent trend is observed below a critical temperature of $85\pm5\,$ K, defined as $T^*$, which is very close to the temperature where the magnetic resistivity exhibits a coherence peak (see the Supplemental Material) \cite{B. Shen2020}. This concurrence indicates that the quasiparticle relaxation is influenced by the coherent heavy electron state emerging below $T^*$, the same as that have been observed in CeCoIn$_5$.

\begin{figure}
	\includegraphics[width=8.7cm]{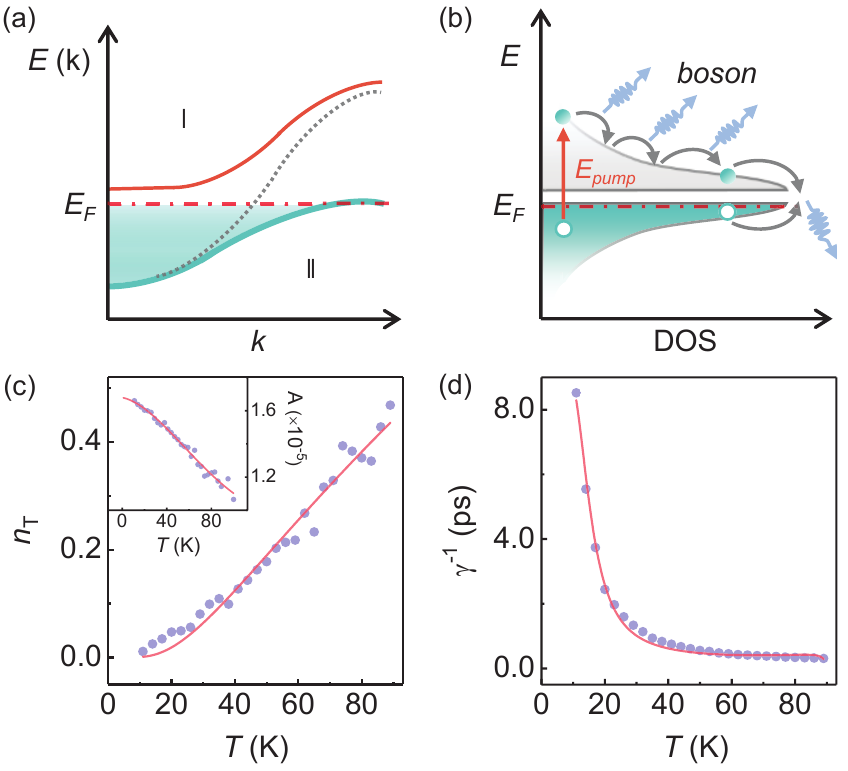}
	\caption{\label{Fig2} (a) Diagram of the collective hybridization between local $f$ and conduction electrons below $T^*$, resulting in the formation of an indirect gap, $2\Delta$. (b) The DOS near the Fermi energy for a Kondo lattice below $T^*$. Nonequilibrium quasparticles are excited with an energy much larger than $2\Delta$. These Quasiparticles decay to the gap edge via emission of high frequency bosons, i.e. phonons or other bosonic excitations. Subsequently, the bimolecular recombination dominates the decay and the relaxation rates become fluence dependent. (c) The density of thermally excited quasiparticles $n_T$ as a function of temperature below $T^*$. The inset shows the temperature dependence of the amplitude $A$. (d) Decay time $\gamma^{-1}$ as a function of temperature below $T^*$. Values of $A$ and $\gamma$ here are for pump fluence of 0.8~$\mu$J/cm$^2$. The red curves are the fitted results using the RT model.}
	\vspace*{-0.4cm}
\end{figure}

The fluence-dependent phenomena of the decay rate $\gamma$ below $T^*$ can be explained by the widely used Rothwarf-Taylor (RT) model \cite{A. Rothwarf1967, V. V. Kabanov2005}, where the dynamics of quasiparticles ($n$) and bosons ($N$) are well described by the presence of a narrow energy gap $ \Delta $ in DOS. Specifically, the relaxation of the excited quasiparticles with an energy larger than the gap ($ \hbar \omega>2\Delta $) is accomplished by the emission of high frequency bosons that can subsequently re-excite electron-hole pairs, as formulated by \cite{A. Rothwarf1967,V. V. Kabanov2005}
\begin{eqnarray}
\frac{dn}{dt}&=&I_0+\beta N-Rn^2,\nonumber\\
\frac{dN}{dt}&=&\frac12[Rn^2-\beta N]-\left(N-N_T\right)\tau_\gamma^{-1},
\end{eqnarray}
where $I_0$ is the external excitation, $n$ is the total number of quasiparticles, $R$ is the recombination rate of electron-hole pairs, $N$ is the density of high frequency bosons with the energy larger than 2$\Delta $, $\beta$ is the probability per unit time for generating the nonequilibrium quasiparticles by such bosons, $\tau_\gamma^{-1}$ is the escaping rate of the high frequency bosons, and $N_T$ is the thermal-equilibrium boson density. As long as $R$ or $\tau_\gamma^{-1}$ is large enough \cite{V. V. Kabanov2005}, the quasiparticle relaxation dynamics is dominated by the so-called bimolecular recombination process, which contributes the nonlinear $n^2$ term and is fluence-dependent. This type of process is schematically shown in Figs.~\ref{Fig2}(a)-(b), and can exactly elucidate the physics behind the fluence-dependent $\gamma$ in Fig.~\ref{Fig1}(c).

The gap formation in the DOS can be studied by fitting the $\gamma(T)$ and $n(T)$ using the RT model \cite{J. Demsar_JPCM2006, V. V. Kabanov1999, E. E. M. Chia2007},
\begin{eqnarray}
\gamma(T)&\propto&\left[\frac{\delta}{\zeta n_\text{T}+1}+2n_T\right]\left(\Delta+\alpha T\Delta^4\right),\nonumber\\
n_\text{T}(T)&=&\frac{A(0)}{A(T)}-1\propto \left(T\Delta\right)^p\text{e}^{-\Delta/T},
\label{eq:RT-model1}
\end{eqnarray}
where $\alpha$, $\zeta$ and $\delta$ are fitting parameters, respectively. $n_T$ is the density of quasiparticles thermally excited across the gap and $p$ ($0<p<1$) is a constant  determined by the shape of the gapped DOS \cite{J. Demsar2006}. For a typical DOS of the Bardeen-Cooper-Schrieffer (BCS) form, we may fix $p=0.5$ and obtain a good fit as shown Figs.~\ref{Fig2}(c) and (d). This yields a gap of the magnitude of $2\Delta\approx12\,$ meV, reflecting the formation of an indirect hybridization gap below $T^*$ due to collective hybridization associated with the emergence of coherent heavy electron states near $E_F$.

For temperatures above $T^{*}$, the fluence-independency clearly indicates the closing of the indirect hybridization gap. However, we notice that the behaviour of $\gamma(T)$ can be further separated into two regimes. For $T>T^{\dagger}$ ($\sim$140 K) , $\gamma$ firstly shows a saturation behavior and then decreases slightly as the temperature increases. Such $T$-dependence cannot be explained by the conventional two-temperature model \cite{Allen_PRL_1987}. Rather, it indicates a nonthermal process, i.e., the relaxation time due to the e-e collisions can be longer than the e-ph relaxation time ($\tau_{e-e}>\tau_{e-ph}$) \cite{C. Gadermaier2010}, or the thermal distribution by e-e scatterings cannot be instantaneously attained, even though the excited fermionic quasiparticles may relax close to $E_F$. Such process may be described by a nonequilibrium model \cite{C. Gadermaier2010}: $1/\tau_{e-ph}=3\hbar\lambda\langle\omega^2\rangle/\left(2\pi k_BT_l\right)$, where $\lambda\langle\omega^2\rangle$ represents the e-ph coupling and $T_l$ is the lattice temperature. If we assume $\gamma=1/\tau_{e-ph}$ in the high temperature regime, i.e. $>$230~K, this model can well explain why the measured $\gamma$ becomes smaller as $T$ increases (see Fig.~\ref{Fig1}(d)), which strongly supports the existence of the nonthermal process within the initial several picoseconds. We further obtain $\lambda\langle\omega^2\rangle\simeq$94~meV$^2$. Assuming the Debye frequency to be 7~THz (or 28~meV) based on our measurements discussed below, we can estimate that $\lambda$ takes a value of $\sim$0.23.

\begin{figure}
	\includegraphics[width=8.9cm]{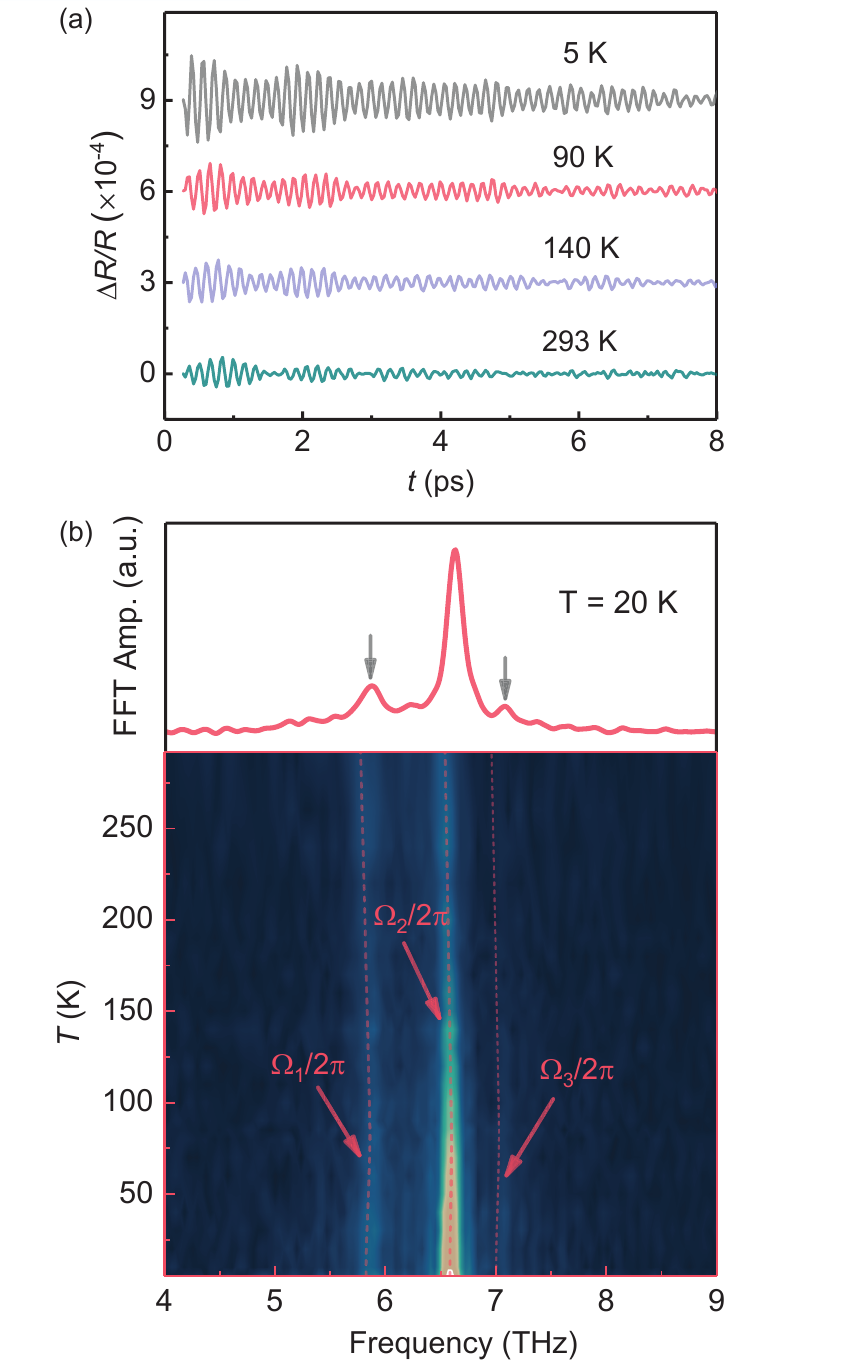}
	\caption{\label{Fig3} (a) Extracted oscillatory components at several typical temperatures, e.g. 5 K, 90 K, 140 K and 293 K. (b) The Fourier transform spectra in the frequency domain for the extracted oscillations from 260 K down to 5 K. Three coherent modes are indicated by the dashed lines.}
	\vspace*{-0.2cm}
\end{figure}

\begin{figure}
	\includegraphics[width=8.2cm]{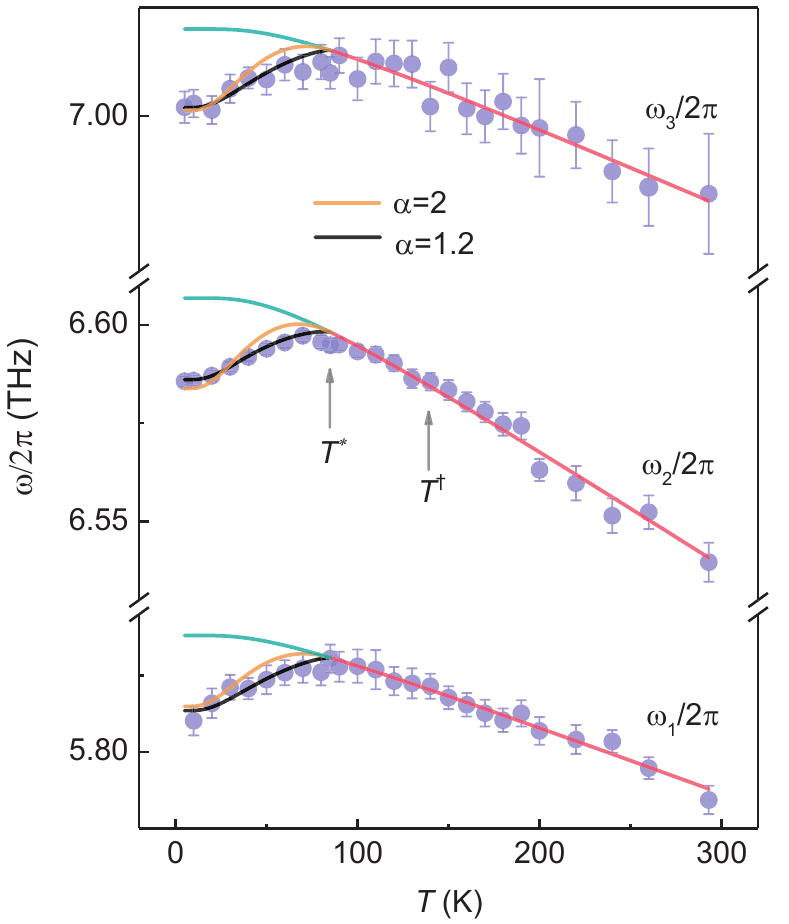}
	\caption{\label{Fig4}The derived $\omega_j$ ($j=1,2,3$) as a function of temperature using Eq.~(3). The green lines represent the fit using the anharmonic phonon model. The orange and black lines are the fit taking into consideration the contribution of Kondo singlets with different $\alpha$, as described in the main text.}
	\vspace*{-0.4cm}
\end{figure}

Between $T^*$ and $T^\dagger$, $\gamma(T)$ decreases with lowering temperature. Similar $T$-dependence has been observed in CeCoIn$_5$ \cite{Y. P. Liu2020}, where, in the absence of the indirect hybridization gap, precursor hybridization fluctuations in a short correlation time- or length-scale were proposed to explain the reduction of decay rate $\gamma$ below $T^\dagger$ \cite{D. Hu2019}. In specific, both the e-e and electron-boson scatterings of the excited quasiparticles are expected to be suppressed as the fluctuating $f$ moments start to participate in the hybridization with the conduction electrons. The development of hybridization fluctuations can further cause the renormalization (bending) of the conduction bands even in the high temperature regime (e.g. $>$100~K) \cite{Q. Y. Chen2017}, a prediction based on our experimental observation to be examined using ARPES in CeRh$_6$Ge$_4$. This suggests that the $f$ moments are not decoupled from conduction electrons in the paramagnetic state as one may naively think for a small Fermi surface at ambient pressure. Note that how the hybridization process evolves below 5 K remains to be investigated.

The above results demonstrate that the hybridization dynamics in CeRh$_6$Ge$_4$ also exhibits a two-stage process, with the onset of precursor hybridization fluctuations below $T^\dagger$ and the opening of an indirect hybridization gap below $T^*$, in resemblance of that in CeCoIn$_5$. The separation of the two stages is also manifested by the oscillatory component superimposed on the $\Delta R(t)/R$ signals. The oscillations with terahertz (THz) frequency generally originate from the coherent optical phonons triggered by displacive excitations or photoexcited Raman process \cite{H. J. Zeiger1992, R. Merlin1997}.

Figures~\ref{Fig3}(a) and (b) plot the oscillatory components extracted from the decay process after subtracting the nonoscillatory background. Three obvious terahertz modes were observed at all investigated temperatures including $\Omega_1/2\pi\sim$5.8 THz, $\Omega_2/2\pi\sim$6.6 THz, $\Omega_3/2\pi\sim$7 THz. The oscillation components can be fitted quantitatively using the following formula,
\begin{equation}
	(\Delta R/R)_{\text{osc}}=\sum_{j=1,2,3}A_je^{-\Gamma_j t}\text{sin}(\Omega_j t+\phi_j),
	\\
	\label{eq:oscillation1}
\end{equation}
where  $A_j$, $\Gamma_j$, $\Omega_j$, and $\phi_j$ ($j=1,2,3$) are the amplitude, damping rate, frequency, and phase, respectively. $\Omega_j$ and $\phi_j$ are related to an underdamped harmonic oscillator. $\Omega_j=\sqrt{\omega_j^2-\Gamma_j^2}$, where $\omega_j$ is the natural frequency. The temperature-dependent evolution of $\omega_j$ is shown in Fig. 4. We see a sharp downturn below $T^*$ for all three phonon modes, instead of a gradual flattening expected by the anharmonic decay model \cite{M. Balkanski1983, J. Menendez1984} (see also the Supplemental Material), as indicated by the green lines in Fig.~\ref{Fig4}. By contrast, the $T$-dependent $\omega_j$ for $T>T^*$ can be well explained by the anharmonic phonon-phonon coupling  (red lines).

The anomaly around $T^*$ cannot arise from the phonon-magnon coupling effect because the long-range FM ordering in CeRh$_6$Ge$_4$ appears below a Curie temperature of 2.5 K. To understand quantitatively the peculiar behavior of $\omega_j(T)$ for $T<T^*$, we calculated the values of $\delta\omega_j$ that represent the deviation between the expected values from anharmonic phonon model and the experimental $\omega_j/2\pi$. Such deviation must be associated with the appearance of collective hybridization and the consequent  gap opening in the DOS. In fact, $\delta\omega_j$ can be connected to the quasiparticle density ($n_T$) via the density of Kondo singlets $\langle b_i\rangle$ \cite{Y. P. Liu2020}, which is proportional to $[1-n_T(T)/n_T(T^*)]$. Our best fit using $\delta\omega_j\propto\langle b_i\rangle^\alpha$, where $\alpha$ is a fitting parameter, yields $ \alpha=1.2\pm0.16$, as shown in Fig.~\ref{Fig4}. Surprisingly, this value of $\alpha$ is nearly the same as that obtained in CeCoIn$_5$ within the experimental errors, while the mean-field theory predicted $\alpha=2$. This indicates that a new and generic theory is required in order to explain the anomalous phonon softening below $T^*$ in two quite different systems. In essence, this theory should take into account variation of the electron-phonon coupling induced by the DOS change near $E_F$ in the presence of collective hybridization. Clearly, the observed frequency softening further proves that the phonon renormalization can be a useful probe of the coherent heavy electron states.

Altogether, our results reveal the detailed hybridization process in the FM Kondo lattice compound CeRh$_6$Ge$_4$ over a wide temperature range. The reduction of the relaxation rate below $T^\dagger\sim$140 K suggests the possibility of ``band bending" already in this high temperature region, which could be detected in future ARPES experiments. Below the coherence temperature $T^*\sim$85~K, we unveil an indirect band gap of $\sim$12~meV, which plays the role of a protector of the coherent heavy electron state. The associated anomalous softenning in the frequencies of coherent optical phonons provides a benchmark for further theories. The observed two-stage hybridization process is in close resemblance of that in CeCoIn$_5$. The distinction in the magnetic correlations appear to have no significant influence at least in the measured temperature range (above 5 K) and at ambient pressure. This seems to suggest a universal mechanism for the onset of heavy electron coherence independent of the details of the inter-site coupling among localized $f$ moments. We should also note that our results in the normal state are in contrast with the observation of a small Fermi surface in the ferromagnetic ground state \cite{A. Wang2021}. This distinction was not anticipated in previous theories. It remains open how the hybridization process changes upon entering the magnetic state. More elaborated studies are needed to clarify this point, and particularly its influence on the properties of quantum criticality \cite{Wang_arXiv_2020}.

\textbf{Note:} During the submission of our manuscript, we realize an ARPES work on CeRh$_6$Ge$_4$ by Y. Wu et al. \cite{Wu_submitted}, providing evidence for anisotropic hybridization between $f$- and conduction electrons in the high temperature regime.

This work was supported by the National Natural Science Foundation of China (Grants No. 11974070, No. 11974306, No. 12034017, No. 11734006, No. 11774401, and No. 11974306), the Frontier Science Project of Dongguan (2019622101004), the National Key R$ \& $D Program of China (Grants No. 2017YFA0303100, No. 2018YFA0307400, and No. 2016YFA0300202), the Science Challenge Project of China (No. TZ2016004), the Chinese Academy of Sciences Interdisciplinary Innovation Team, the Strategic Priority Research Program of CAS (XDB33010100), and the Key R\&D Program of Zhejiang Province, China (2021C01002).

\end{document}